\documentclass[
superscriptaddress,
preprintnumbers,
preprint,
12pt
]{revtex4-1}
\usepackage{psfrag}

\usepackage{amsmath}
\usepackage{amssymb}

\usepackage{fullpage}

\usepackage{graphicx}

\begin{document}

\title{Warped 5D Standard Model Consistent with EWPT}

\author{Joan A. Cabrer}
%
%\email{}
%\homepage{}
\affiliation{%
Institut de F\'isica d'Altes Energies,\\ Universitat Aut{\`o}noma de Barcelona, 08193 Bellaterra, Barcelona, Spain\nobreak
\vspace{3mm}
}%

%\collaboration{}%\noaffiliation

\author{Gero von Gersdorff}
%\email{}
\affiliation{
Centre de Physique Th\'eorique, \'Ecole Polytechnique and CNRS, F-91128 Palaiseau, France
\vspace{3mm}
}%

\author{Mariano Quir\'os}
%\email{quiros@ifae.es}
\affiliation{
Instituci\'o Catalana de Recerca i Estudis  
Avan\c{c}ats (ICREA) and \nobreak
\\
Institut de F\'isica d'Altes Energies,
\\
Universitat Aut{\`o}noma de Barcelona, 08193 Bellaterra, Barcelona, Spain\nobreak
\vspace{1.7cm}
\\
{\normalfont Based on talk given by M. Quir\'os at the \\ {\it Workshop on the Standard Model and Beyond -- Cosmology},\\ Corfu Summer Institute, Greece, August 29 -- September 5, 2010}
\vspace{1.7cm}
\\
}%

\begin{abstract}
For a 5D Standard Model propagating in an AdS background with an IR localized Higgs, compatibility of bulk KK gauge modes with EWPT yields a phenomenologically unappealing KK spectrum ($m_{KK}\geq 12.5$ TeV)  and leads to a ``little hierarchy problem''. For a bulk Higgs the solution to the hierarchy problem reduces the previous bound only by $\sim\sqrt{3}$. As a way out,  models with an enhanced bulk gauge symmetry $SU(2)_R\times U(1)_{B-L}$ were proposed. In this note we describe a much simpler (5D Standard) Model, where introduction of an enlarged gauge symmetry is no longer required. It is  based on a warped gravitational background which departs from AdS at the IR brane and a bulk propagating Higgs. The model is consistent with EWPT for a range of KK masses within the LHC reach.
\end{abstract}
%% maketitle must follow the abstract.
\maketitle                   % Produces the title.

%% If there is not enough space inside the running head
%% for all authors including the title you may provide
%% the leftmark in one of the following three forms:

%% \renewcommand{\leftmark}
%% {First Author: A Short Title}

%% \renewcommand{\leftmark}
%% {First Author and Second Author: A Short Title}

%% \renewcommand{\leftmark}
%% {First Author et al.: A Short Title}

%% \tableofcontents  % Produces the table of contents.

%\section{Introduction}

Warped models in extra dimensions with two boundaries (the UV and IR branes), were proposed by Randall and Sundrum (RS)~\cite{Randall:1999ee} as a very elegant solution to the hierarchy problem. Provided the Higgs is localized towards the IR boundary, its mass is indeed redshifted from the Planck to the TeV scale  in the four-dimensional (4D) theory, thus removing the UV sensitivity of the Standard Model (SM) Higgs mass, known as the hierarchy problem. 

A general warped 5D metric can be written as
$
ds^2=e^{-2A(y)}\eta_{\mu\nu}dx^\mu dx^\nu+dy^2\,
%\label{metrica}
$
in proper coordinates, where $A(y)=k y$ in the RS model, $k$ being the AdS curvature constant (of the order the Planck scale) and with the UV (IR) boundary located at $y=0$ ($y=y_1$). If the Higgs profile behaves as
$
h(y)=h(0)e^{a k y}
$,
the AdS/CFT correspondence yields the dimension of the Higgs condensate as ${\rm dim}(\mathcal O_H)=a$. The hierarchy problem is thus solved when $a>2$~\cite{Luty:2004ye}. In the RS background the KK gauge bosons have a profile given by
\begin{equation}
f_n(y)=\mathcal N_nz(y)\left(Y_0\left[m_n/k\right]J_1\left[m_n z(y)\right]-J_0\left[m_n/k\right]Y_1\left[m_n z(y)\right] \right)
\,,
\label{KK-RS}
\end{equation}
where $z(y)=e^{ky}/k$ is the conformally flat coordinate and the mass eigenvalues are defined by $m_n= \nobreak j_{0,n}\rho$, where $j_{0,n}$ is $n^{\rm th}$ zero of the Bessel function $J_0(x)$ and $\rho=ke^{-ky_1}$. When KK gauge bosons propagate in the 5D bulk they contribute to the electroweak precision observables (EWPO), in particular to the $T$ and $S$ parameters~\cite{Peskin:1991sw}, and their masses and couplings have to be contrasted with the electroweak presicion tests (EWPT) which translate into lower bounds on their masses. In particular in RS with a bulk Higgs the prediction for these observables is given by\cite{Round:2010kj,Cabrer:2010si}
\begin{equation}
\alpha(m_Z) T= s^2_W \frac{m_Z^2}{\rho^2}(ky_1)
\frac{(a-1)^2}{a(2a-1)}
,\qquad
\alpha(m_Z)S=2 s_W^2c_W^2
\frac{m_Z^2}{\rho^2}\frac{a^2-1}{a^2}.
\label{SRS}
\end{equation}
One can see from (\ref{SRS}) that the contribution to the $T$ parameter is enhanced by a factor $ky_1\sim 35$ while that of the $S$ parameter is not. This translates into a very strong bound on $m_1\equiv m_{KK}$ when we compare these expressions with the experimental data. A SM fit with a reference Higgs mass of $117~\mathrm{GeV}$ and assuming $U=0$~\cite{Nakamura:2010zzi}, yield
$
T = 0.07 \pm 0.08,\  S = 0.03 \pm 0.09 
$
with a correlation between $S$ and $T$ of $87\%$ in the fit. In particular we find for $m_H=115$ GeV and a localized Higgs ($a\gg 1$) the 95\% CL lower bound $m_{KK}\geq 12.5$ GeV while for a less localized Higgs with $a=2.1$, still consistent with the solution to the hierarchy problem, we find $m_{KK}\geq 7$ GeV.

These bounds make this theory phenomenologically unappealing, since they are much larger than LHC scales and create a ``little hierarchy problem" which translates into some amount of fine-tuning to stabilize weak masses. In order to avoid the large volume-enhanced contributions to the $T$ parameter it was proposed to enlarge the gauge symmetry in the bulk by adding the $SU(2)_R\times U(1)_{B-L}$ gauge group~\cite{Agashe:2003zs} such that only the large contributions to the $S$ parameter should be taken care which yields bounds on KK masses of $\mathcal O(3)$ TeV. In this note we will present a 5D SM propagating in the bulk consistent with EWPT for KK-masses in the LHC range based on a gravitational background which is nearly AdS at the UV boundary but departs from it at the IR brane.

In the rest of this note we will explore an alternative solution to the problem of the $T$ parameter based on a general 5D metric which behaves as AdS near the UV boundary and departs from it near the IR brane. We will see that the combined effect of the Higgs delocalization and the departure from AdS at the IR  region will make the $T$ and the $S$ parameters comparable to each other thus providing a "gravitational solution" to the $T$ problem without any need to introduce and gauge an extra custodial symmetry.

\begin{figure}[h]\begin{center}
\begin{psfrags}
\input{a-psfrag.tex}
\includegraphics[height=8.0cm]{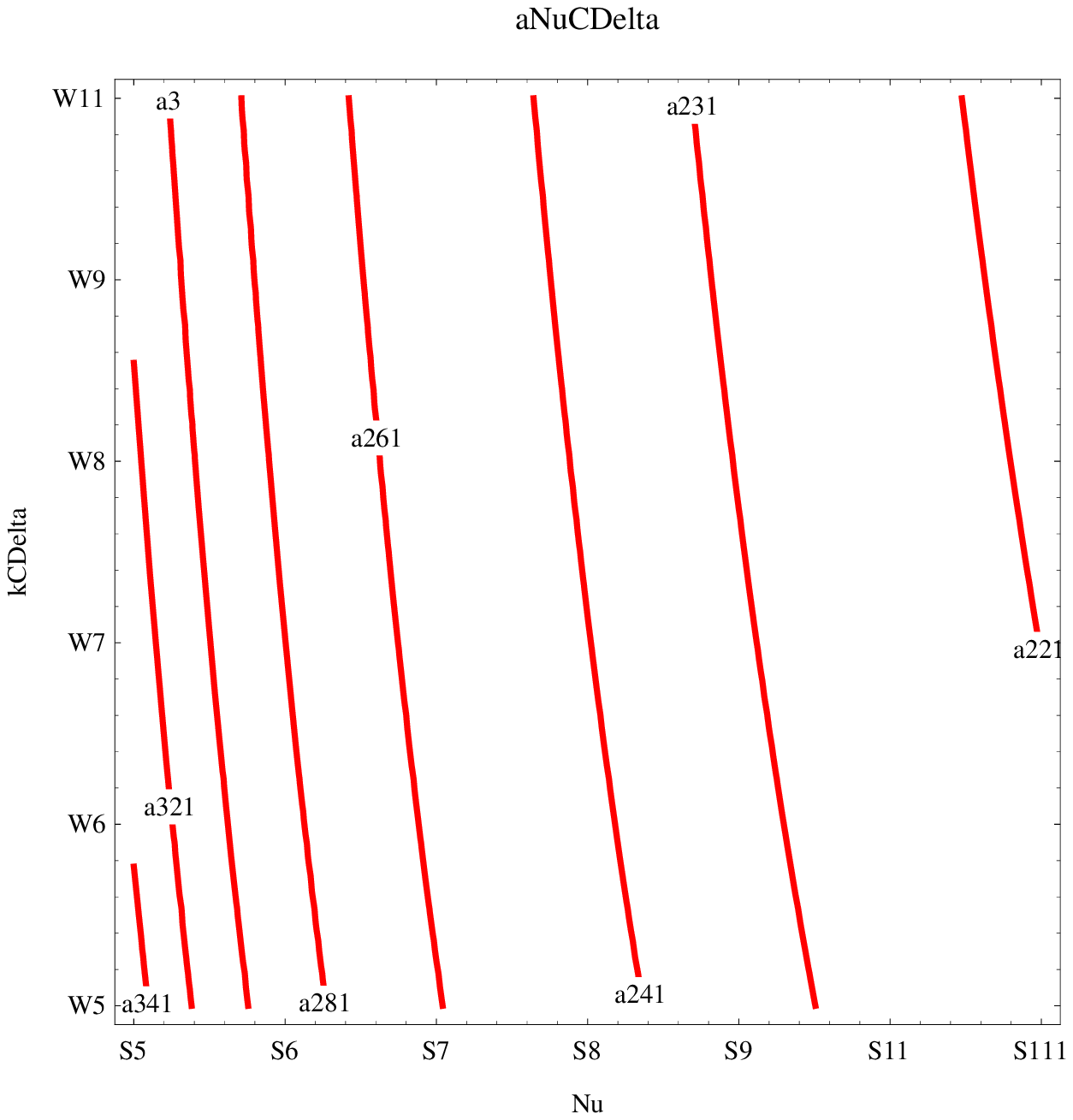}
\end{psfrags}
\hspace{.5cm}
\begin{psfrags}
\input{y1-psfrag.tex}
\includegraphics[height=8.0cm]{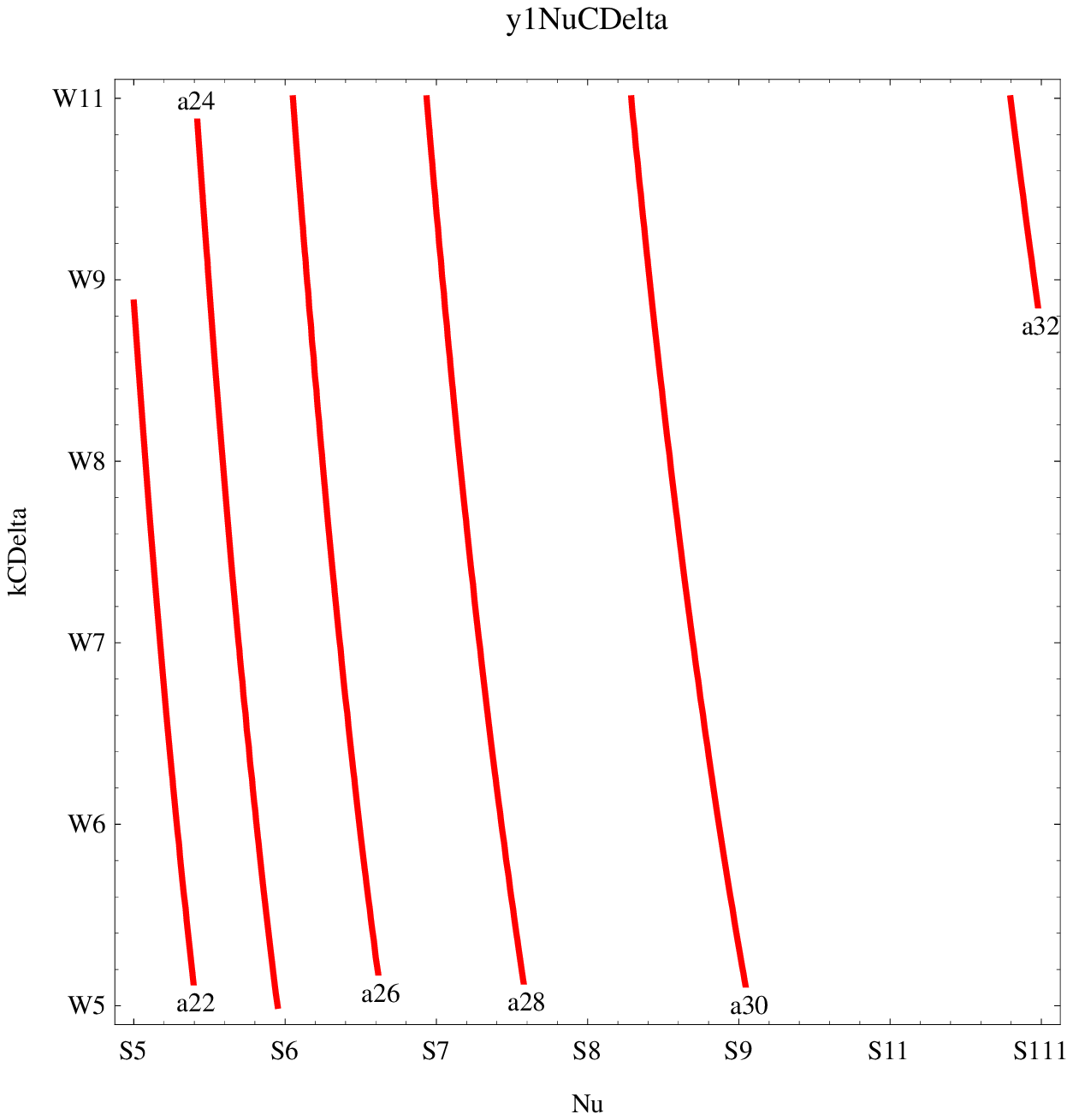}
\end{psfrags}
\caption{Left panel: Contour lines of fixed $a_0(\nu,\Delta)$ from the solution of the hierarchy problem. Right panel: Contour lines of fixed $y_1(\nu,\Delta)$ fixing $A(y_1)=35$.}
\label{fig:1}\end{center}
\end{figure}

We will consider the metric~\cite{Cabrer:2009we}
\begin{equation}
A(y)=ky-\frac{1}{\nu^2}\log\left(1-\frac{y}{y_s}\right) \,,
\label{nuestrametrica}
\end{equation}
where $\nu$ is a real parameter and $y_s=y_1+\Delta$ is the location of a curvature singularity at a distance $\Delta$ from the IR brane, outside the physical interval. 
A dynamical model for the metric Eq.~(\ref{nuestrametrica}) has been described in Refs.~\cite{Cabrer:2009we,Cabrer:2010si}.
Departure from AdS is provided by finite values of $\nu$ and $\Delta$. The KK-gauge bosons propagating in the background (\ref{nuestrametrica}) have profiles given by Eq.~(\ref{KK-RS}) where the conformally flat coordinate can be approximated by $z(y)\approx e^{A(y)}/A'(y)$ and the mass eigenvalues by $m_n\approx j_{0,n}A'(y_1)\rho/k$ where $\rho=ke^{-A(y_1)}$~\cite{Cabrer:2011fb} and $A(y_1)\sim 35$ to solve the hierarchy problem. The latter condition fixes the volume $ky_1<35$ in terms of the remaining parameters. A suitable bulk Higgs mass leads to $h(y)=c_1 e^{a ky}+c_2 \int^ye^{4A(y')-2aky'}$, and $h(y)\sim e^{aky}$ imposes the constraint $a>a_0=2 A_1/ky_1$ as we analyzed in Ref.~\cite{Cabrer:2011fb}.
Contour lines for the bound $a_0(\nu,\Delta)$ are shown in Fig.~\ref{fig:1} (left panel). We can see there that the main dependence of $a_0$ is on $\nu$ and that values of $a$ much larger than 2 can be required to solve the hierarchy problem. In the RS limit $a_0\to 2$.

\begin{figure}[h]\begin{center}
\begin{psfrags}
\input{Z-psfrag.tex}
\includegraphics[height=8.0cm]{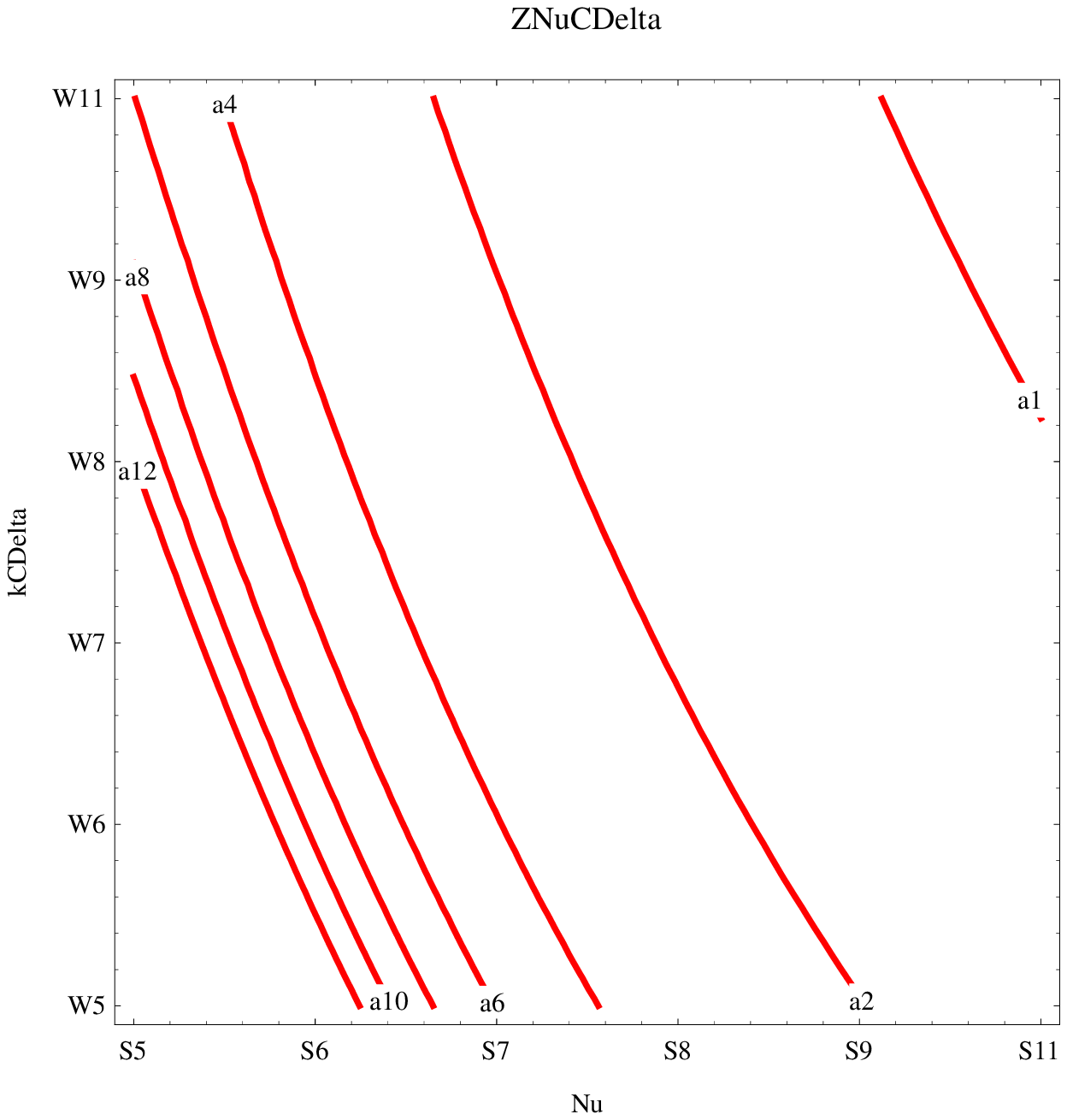}
\end{psfrags}
\hspace{.5cm}
\begin{psfrags}
\input{TsobreS-psfrag.tex}
\includegraphics[height=8.0cm]{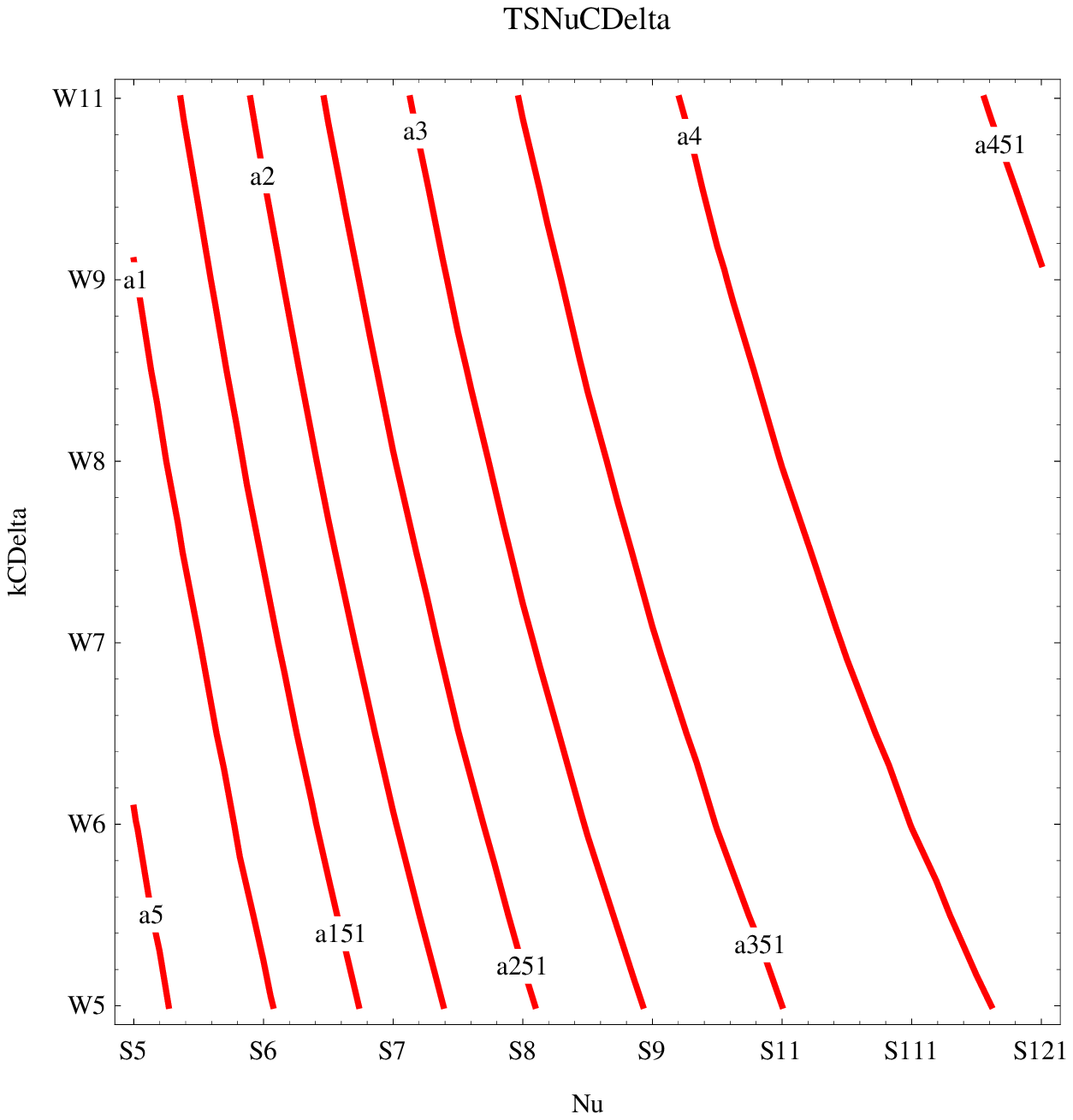}
\end{psfrags}
\caption{Left panel: Contour lines of fixed $Z(\nu,\Delta)$ where $a=a_0(\nu,\Delta)$ and $A(y_1)=35$. Right panel: The same as in the left panel but for $T/S$.}
\label{fig:2}\end{center}
\end{figure}

\begin{figure}[h]\begin{center}
\begin{psfrags}
\input{bounds-psfrag.tex}
\includegraphics[height=8.0cm]{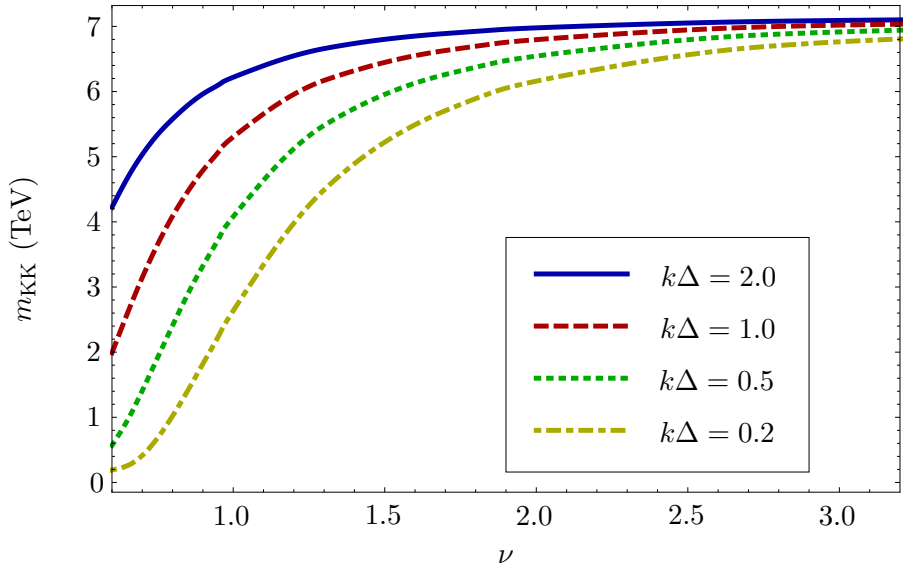}
\end{psfrags}
\caption{95\%~C.L. bounds on the mass of the first KK mode of the gauge boson as a function of $\nu$ and for different values of $\Delta$, taking into account only the contribution from $S$ and $T$ parameters. For each curve we have set $a=a_0(\nu,\Delta)$.}
\label{fig:bounds}\end{center}
\end{figure}

The prediction for the observables $T$ and $S$ is given by~\cite{Cabrer:2010si,Cabrer:2011fb}
\begin{eqnarray}
\alpha T&=&s_W^2m_Z^2y_1\int_0^{y_1} e^{2A}(1-\Omega)^2
= s_W^2 \frac{m_Z^2}{\rho^2}\, I_2\, \frac{ky_1}{Z^2}\,\,,\nonumber\\
\alpha S&=&8s_W^2c_W^2 m_Z^2 \int_0^{y_1} e^{2A}\left(y_1-y\right)(1-\Omega)
= 8 s^2_W c^2_W \frac{m_Z^2}{\rho^2}\, I_1\,\,\frac{1}{Z} \,,
\label{ST}
\end{eqnarray}
where $1-\Omega(y)=k\,u(y)/Z,\quad u(y)=\int_y^{y_1}\,e^{-2A(y')+2 A(y_1)}h^2(y')/h^2(y_1)$ and
\begin{equation}
Z=
k\int_0^{y_1}\frac{h^2(y)}{h^2(y_1)}e^{-2A(y)+2A(y_1)}\,,\
I_n=k^3\int_0^{y_1} (y_1-y)^{2-n}u^n(y)e^{2A(y)-2A(y_1)}
\label{Z}
\end{equation}
$Z$ is an additional wave function renormalization depending on both the 
gravitational and Higgs backgrounds and functions $I_{n}/\rho^2$ are $\mathcal O(1/m_{KK}^2)$ and not very sensitive to the model parameters as it was shown in Ref.~\cite{Cabrer:2011fb}. We can see from the expressions in (\ref{ST}) that $T$ is indeed volume enhanced while $S$ is not. As discussed above $ky_1<A(y_1)$ and hence the volume will be reduced a little due to them deformation, which will suppress a little bit the $T$-parameter. We can see this effect in Fig.~\ref{fig:1} (right panel) where we can see that this effect is moderate and depends mainly on the $\nu$ parameter. In the RS limit one finds $ky_1\to A(y_1)$.
The main reason for the reduction in $T$ and $S$ thus comes from the Higgs renormalization $Z$.
 In the left panel of Fig.~\ref{fig:2}  we can see that depending on the regions in the $(\nu,\Delta)$ plane the factor $Z$ can be $Z\gg 1$. In fact, in the right panel of Fig.~\ref{fig:2} we can see that correspondingly in those regions of the $(\nu,\Delta)$ plane the observable $T$ can be of the order  of (or even smaller than) the $S$ observable.
In Fig.~\ref{fig:bounds} we show the 95\%~C.L. bounds on $m_{KK}$ as a function of $\nu$ and different values of $\Delta$. In this plot we take $a=a_0(\nu,\Delta)$, the minimum value of $a$ consistent with solving the hierarchy problem. We can see from this figure that, depending on the value of $k\Delta$, for small values of $\nu$ we can achieve bounds $m_{KK} \gtrsim \mathcal{O}(1)~\mathrm{TeV}$ or even lower (which would have be to be contrasted with experimental direct detection bounds). For large values of $\nu$ the bounds converge to the RS result ($m_{KK} > 7~\mathrm{TeV}$).

Finally we will conclude this note with a couple of short comments on the possible limits on the model parameters $(\nu,\Delta)$. As we can see from Fig.~\ref{fig:bounds} the lower bounds on $m_{KK}$ can apparently go much below 1~TeV for values of $\nu$ and/or $\Delta\ll 1$. However in this case there are other oblique observables that will become important and cannot be neglected, corresponding to four-fermion effective operators, in particular the $W$ and $Y$ observables~\cite{Barbieri:2004qk} which in our model are given by
\begin{equation}
W=Y=\frac{c_W^2m_Z^{2}}{y_1}
\int e^{2A}\left(y_1-y\right)^2
=c_W^2 m_Z^2\, \frac{I_0}{\rho^2}\,\frac{1}{ky_1}\,.
\end{equation}
It turns out that for $m_{KK}\ll 1$ TeV the $W$ and $Y$ observables will be the leading ones and they can exceed their corresponding experimental values~\cite{Barbieri:2004qk}. A second comment is that in the region where $k\nu^2\Delta<1$ the curvature at $y=y_1$ can become large (due to the proximity of the spurious singularity at $y=y_1+\Delta$). Correspondingly the curvature radius $L_1\simeq \nu^2\Delta$ decreases, apparently jeopardizing perturbativity in the 5D gravity theory which is controlled by the parameter $M_5L_1$. However the latter remains under control at least for $kL_1>0.2$ as we have proven in Ref.~\cite{Cabrer:2011fb} because $M_5L_1=(M_5/k) kL_1>1$, essentially because $k\ll M_5$ thus leading to $M_5L_1> 1$.

\newpage
{
{\bf Acknowledgements~~} Work supported in part by the Spanish Consolider-Ingenio 2010
Programme CPAN (CSD2007-00042) and by CICYT-FEDER-FPA2008-01430.  The work of JAC is supported by the Spanish Ministry of Education through a FPU grant. The research of GG is supported by the ERC Advanced Grant 226371, the ITN programme PITN- GA-2009-237920 and the IFCPAR CEFIPRA programme 4104-2.   
}

\end{document}